\documentstyle[sprocl]{article}

\def\Journal#1#2#3#4{{#1} {\bf #2}, #3 (#4)}

\def\NPB{{\em Nucl. Phys.} B}
\def\PLB{{\em Phys. Lett.}  B}

\def\PRD{{\em Phys. Rev.} D}

\def\AP{\em Ann. Phys. (N.Y.)}
\def\PR{\em Phys. Rev.}

\def\be{\begin{equation}}
\def\ee{\end{equation}}
\def\bea{\begin{eqnarray}}
\def\eea{\end{eqnarray}}
\def\bfg{\begin{figure}[ht]}
\def\efg{\end{figure}}  
\def\lb{\label}

\begin{document}

\title{RELATIONSHIP OF PIONIUM LIFETIME WITH PION SCATTERING LENGTHS
IN GENERALIZED CHIRAL PERTURBATION THEORY}  

\author{H. SAZDJIAN}

\address{Groupe de Physique Th\'eorique, Institut de Physique Nucl\'eaire,\\
Universit\'e Paris XI, F-91406 Orsay Cedex, France\\
E-mail: sazdjian@ipno.in2p3.fr}


\maketitle\abstracts{The pionium lifetime is calculated in the framework
of the quasipotential-constraint theory approach, including the sizable
electromagnetic corrections. The framework of generalized chiral
perturbation theory allows then an analysis of the lifetime value as a
function of the $\pi\pi$ $S$-wave scattering lengths with isospin $I=0,2$,
the latter being dependent on the quark condensate value.}

\noindent 

The DIRAC experiment at CERN is expected to measure the pionium
lifetime with a 10\% accuracy. The pionium is an atom made of $\pi^+\pi^-$,
which decays under the effect of strong interactions into $\pi^0\pi^0$.
The physical interest of the lifetime is that it gives us information
about the $\pi\pi$ scattering lengths. The nonrelativistic formula of the 
lifetime was first obtained by Deser {\it et al.} \cite{dgbt}:
\begin{equation} \label{e1}
\frac{1}{\tau_0} = \Gamma_0 = \frac{16\pi}{9}\sqrt
{\frac{2\Delta m_{\pi}}
{m_{\pi^+}}} \frac{(a_0^0-a_0^2)^2}{m_{\pi^+}^2} |\psi_{+-}(0)|^2,\ \ \
\ \ \Delta m_{\pi}=m_{\pi^+}-m_{\pi^0},
\end{equation}
where $\psi_{+-}(0)$ is the wave function of the pionium at the origin
(in $x$-space) and $a_0^0$, $a_0^2$, the $S$-wave scattering lengths with
isospin 0 and 2, respectively.
\par
The evaluation of the relativistic corrections to this formula can be
done in a systematic way in the framework of chiral perurbation theory
($\chi PT$) \cite{gl}, in the presence of electromagnetism \cite{uk}. 
There arise essentially two types of correction. (i) The pion-photon 
radiative corrections, which are similar to those met in conventional QED.
(ii) The quark-photon radiative corrections, which appear through terms
where the photon field is not explicitly present and which are mainly 
responsible for the pion mass difference at lowest-order. 
\par
Three different methods of evaluation have been used for the study of the
pionium bound state in the framework of $\chi PT$. The first method uses a 
three-dimensionally reduced form of the Bethe--Salpeter equation within
the quasi\-poten\-tial--constraint theory approach  \cite{jss}. The second
method uses the Bethe--Salpeter equation with the Coulomb gauge \cite{illr}.
The third one uses the approach of nonrelativistic effective theory 
\cite{gglr}. All the above approaches lead to
similar estimates, on the order of $6\%$, for the relativistic corrections
to the nonrelativistic formula of the pionium decay width. 
\par
The theoretical interest of the $\pi\pi$ scattering lengths is that they
allow us to estimate the value of the quark condensate in QCD. The 
fundamental order parameter of spontaneous chiral symmetry breaking being
$F_{\pi}$, the pion weak decay constant, other order parameters may 
eventually vanish in the chiral limit without contradicting chiral symmetry
breaking, as long as $F_{\pi}$ remains different from zero in that limit.
Such an issue is intimately dependent on the mechanism of chiral symmetry 
breaking. In standard $\chi PT$ \cite{gl}, it is assumed that the quark 
condensate parameter, defined as $<0|\overline qq|0>/F_{\pi}^2$, is on the
order of the hadronic mass scale ($\sim 1$ GeV). This hypothesis is verified 
in the sigma-model and the Nambu--Jona-Lasiono model. The vacuum state here
is similar to a ferromagnetic type medium. On the other hand, in an 
antiferromagnetic type medium, one would have a vanishing quark condensate 
and yet chiral symmetry would still be broken \cite{l,s}. An intermediate 
possibility, due to an eventual phase transition in QCD for large values of
the light quark flavor number, was also advocated recently \cite{dgs}.
\par
Generalized $\chi PT$ is a framework in which the quark condensate value is
left as a free parameter subjected to an experimental evaluation \cite{fss}.
The Goldstone boson scattering amplitudes are sensitive to the quark
condensate value and hence their experimental measurment gives us an
estimate of the latter quantity. Thus, in the $\pi\pi$ scattering amplitude
relatively small values of the $S$-wave isospin-0 scattering length $a_0^0$,
on the order of, say, 0.21-0.22, correspond to the predictions of standard 
$\chi PT$, while relatively large values of $a_0^0$, on the
order of, say, 0.28-0.36, correspond to small values of the quark condensate
parameter. 
\par
We have redone the analysis of the pionium lifetime in the framework of
generalized $\chi PT$ \cite{sz}. Eliminating the quark condensate parameter
in favor of the combination $(a_0^0-a_0^2)$ we have calculated the pionium
lifetime as a function of $(a_0^0-a_0^2)$. The corresponding curve is
presented in Fig. \ref{f1}.
\par
\bfg
\vspace*{0.5 cm}
\begin{center}
\setlength{\unitlength}{0.240900pt}
\ifx\plotpoint\undefined\newsavebox{\plotpoint}\fi
\sbox{\plotpoint}{\rule[-0.200pt]{0.400pt}{0.400pt}}%
\begin{picture}(1350,990)(0,0)
\font\gnuplot=cmr10 at 10pt
\gnuplot
\sbox{\plotpoint}{\rule[-0.200pt]{0.400pt}{0.400pt}}%
\put(181.0,163.0){\rule[-0.200pt]{4.818pt}{0.400pt}}
\put(161,163){\makebox(0,0)[r]{1}}
\put(1310.0,163.0){\rule[-0.200pt]{4.818pt}{0.400pt}}
\put(181.0,294.0){\rule[-0.200pt]{4.818pt}{0.400pt}}
\put(161,294){\makebox(0,0)[r]{1.5}}
\put(1310.0,294.0){\rule[-0.200pt]{4.818pt}{0.400pt}}
\put(181.0,425.0){\rule[-0.200pt]{4.818pt}{0.400pt}}
\put(161,425){\makebox(0,0)[r]{2}}
\put(1310.0,425.0){\rule[-0.200pt]{4.818pt}{0.400pt}}
\put(181.0,556.0){\rule[-0.200pt]{4.818pt}{0.400pt}}
\put(161,556){\makebox(0,0)[r]{2.5}}
\put(1310.0,556.0){\rule[-0.200pt]{4.818pt}{0.400pt}}
\put(181.0,687.0){\rule[-0.200pt]{4.818pt}{0.400pt}}
\put(161,687){\makebox(0,0)[r]{3}}
\put(1310.0,687.0){\rule[-0.200pt]{4.818pt}{0.400pt}}
\put(181.0,818.0){\rule[-0.200pt]{4.818pt}{0.400pt}}
\put(161,818){\makebox(0,0)[r]{3.5}}
\put(1310.0,818.0){\rule[-0.200pt]{4.818pt}{0.400pt}}
\put(181.0,949.0){\rule[-0.200pt]{4.818pt}{0.400pt}}
\put(161,949){\makebox(0,0)[r]{4}}
\put(1310.0,949.0){\rule[-0.200pt]{4.818pt}{0.400pt}}
\put(181.0,163.0){\rule[-0.200pt]{0.400pt}{4.818pt}}
\put(181,122){\makebox(0,0){0.25}}
\put(181.0,929.0){\rule[-0.200pt]{0.400pt}{4.818pt}}
\put(277.0,163.0){\rule[-0.200pt]{0.400pt}{4.818pt}}
\put(277,122){\makebox(0,0){0.26}}
\put(277.0,929.0){\rule[-0.200pt]{0.400pt}{4.818pt}}
\put(373.0,163.0){\rule[-0.200pt]{0.400pt}{4.818pt}}
\put(373,122){\makebox(0,0){0.27}}
\put(373.0,929.0){\rule[-0.200pt]{0.400pt}{4.818pt}}
\put(468.0,163.0){\rule[-0.200pt]{0.400pt}{4.818pt}}
\put(468,122){\makebox(0,0){0.28}}
\put(468.0,929.0){\rule[-0.200pt]{0.400pt}{4.818pt}}
\put(564.0,163.0){\rule[-0.200pt]{0.400pt}{4.818pt}}
\put(564,122){\makebox(0,0){0.29}}
\put(564.0,929.0){\rule[-0.200pt]{0.400pt}{4.818pt}}
\put(660.0,163.0){\rule[-0.200pt]{0.400pt}{4.818pt}}
\put(660,122){\makebox(0,0){0.3}}
\put(660.0,929.0){\rule[-0.200pt]{0.400pt}{4.818pt}}
\put(756.0,163.0){\rule[-0.200pt]{0.400pt}{4.818pt}}
\put(756,122){\makebox(0,0){0.31}}
\put(756.0,929.0){\rule[-0.200pt]{0.400pt}{4.818pt}}
\put(851.0,163.0){\rule[-0.200pt]{0.400pt}{4.818pt}}
\put(851,122){\makebox(0,0){0.32}}
\put(851.0,929.0){\rule[-0.200pt]{0.400pt}{4.818pt}}
\put(947.0,163.0){\rule[-0.200pt]{0.400pt}{4.818pt}}
\put(947,122){\makebox(0,0){0.33}}
\put(947.0,929.0){\rule[-0.200pt]{0.400pt}{4.818pt}}
\put(1043.0,163.0){\rule[-0.200pt]{0.400pt}{4.818pt}}
\put(1043,122){\makebox(0,0){0.34}}
\put(1043.0,929.0){\rule[-0.200pt]{0.400pt}{4.818pt}}
\put(1139.0,163.0){\rule[-0.200pt]{0.400pt}{4.818pt}}
\put(1139,122){\makebox(0,0){0.35}}
\put(1139.0,929.0){\rule[-0.200pt]{0.400pt}{4.818pt}}
\put(1234.0,163.0){\rule[-0.200pt]{0.400pt}{4.818pt}}
\put(1234,122){\makebox(0,0){0.36}}
\put(1234.0,929.0){\rule[-0.200pt]{0.400pt}{4.818pt}}
\put(1330.0,163.0){\rule[-0.200pt]{0.400pt}{4.818pt}}
\put(1330,122){\makebox(0,0){0.37}}
\put(1330.0,929.0){\rule[-0.200pt]{0.400pt}{4.818pt}}
\put(181.0,163.0){\rule[-0.200pt]{276.794pt}{0.400pt}}
\put(1330.0,163.0){\rule[-0.200pt]{0.400pt}{189.347pt}}
\put(181.0,949.0){\rule[-0.200pt]{276.794pt}{0.400pt}}
\put(82,876){\makebox(0,0){$\tau\ (fs)$}}
\put(755,61){\makebox(0,0){$a_0^0-a_0^2$}}
\put(181.0,163.0){\rule[-0.200pt]{0.400pt}{189.347pt}}
\multiput(181.00,739.93)(0.758,-0.488){13}{\rule{0.700pt}{0.117pt}}
\multiput(181.00,740.17)(10.547,-8.000){2}{\rule{0.350pt}{0.400pt}}
\multiput(193.00,731.93)(0.798,-0.485){11}{\rule{0.729pt}{0.117pt}}
\multiput(193.00,732.17)(9.488,-7.000){2}{\rule{0.364pt}{0.400pt}}
\multiput(204.00,724.93)(0.874,-0.485){11}{\rule{0.786pt}{0.117pt}}
\multiput(204.00,725.17)(10.369,-7.000){2}{\rule{0.393pt}{0.400pt}}
\multiput(216.00,717.93)(0.798,-0.485){11}{\rule{0.729pt}{0.117pt}}
\multiput(216.00,718.17)(9.488,-7.000){2}{\rule{0.364pt}{0.400pt}}
\multiput(227.00,710.93)(0.874,-0.485){11}{\rule{0.786pt}{0.117pt}}
\multiput(227.00,711.17)(10.369,-7.000){2}{\rule{0.393pt}{0.400pt}}
\multiput(239.00,703.93)(0.874,-0.485){11}{\rule{0.786pt}{0.117pt}}
\multiput(239.00,704.17)(10.369,-7.000){2}{\rule{0.393pt}{0.400pt}}
\multiput(251.00,696.93)(0.798,-0.485){11}{\rule{0.729pt}{0.117pt}}
\multiput(251.00,697.17)(9.488,-7.000){2}{\rule{0.364pt}{0.400pt}}
\multiput(262.00,689.93)(0.874,-0.485){11}{\rule{0.786pt}{0.117pt}}
\multiput(262.00,690.17)(10.369,-7.000){2}{\rule{0.393pt}{0.400pt}}
\multiput(274.00,682.93)(0.943,-0.482){9}{\rule{0.833pt}{0.116pt}}
\multiput(274.00,683.17)(9.270,-6.000){2}{\rule{0.417pt}{0.400pt}}
\multiput(285.00,676.93)(0.874,-0.485){11}{\rule{0.786pt}{0.117pt}}
\multiput(285.00,677.17)(10.369,-7.000){2}{\rule{0.393pt}{0.400pt}}
\multiput(297.00,669.93)(0.874,-0.485){11}{\rule{0.786pt}{0.117pt}}
\multiput(297.00,670.17)(10.369,-7.000){2}{\rule{0.393pt}{0.400pt}}
\multiput(309.00,662.93)(0.943,-0.482){9}{\rule{0.833pt}{0.116pt}}
\multiput(309.00,663.17)(9.270,-6.000){2}{\rule{0.417pt}{0.400pt}}
\multiput(320.00,656.93)(1.033,-0.482){9}{\rule{0.900pt}{0.116pt}}
\multiput(320.00,657.17)(10.132,-6.000){2}{\rule{0.450pt}{0.400pt}}
\multiput(332.00,650.93)(0.874,-0.485){11}{\rule{0.786pt}{0.117pt}}
\multiput(332.00,651.17)(10.369,-7.000){2}{\rule{0.393pt}{0.400pt}}
\multiput(344.00,643.93)(0.943,-0.482){9}{\rule{0.833pt}{0.116pt}}
\multiput(344.00,644.17)(9.270,-6.000){2}{\rule{0.417pt}{0.400pt}}
\multiput(355.00,637.93)(1.033,-0.482){9}{\rule{0.900pt}{0.116pt}}
\multiput(355.00,638.17)(10.132,-6.000){2}{\rule{0.450pt}{0.400pt}}
\multiput(367.00,631.93)(0.943,-0.482){9}{\rule{0.833pt}{0.116pt}}
\multiput(367.00,632.17)(9.270,-6.000){2}{\rule{0.417pt}{0.400pt}}
\multiput(378.00,625.93)(1.033,-0.482){9}{\rule{0.900pt}{0.116pt}}
\multiput(378.00,626.17)(10.132,-6.000){2}{\rule{0.450pt}{0.400pt}}
\multiput(390.00,619.93)(1.033,-0.482){9}{\rule{0.900pt}{0.116pt}}
\multiput(390.00,620.17)(10.132,-6.000){2}{\rule{0.450pt}{0.400pt}}
\multiput(402.00,613.93)(1.155,-0.477){7}{\rule{0.980pt}{0.115pt}}
\multiput(402.00,614.17)(8.966,-5.000){2}{\rule{0.490pt}{0.400pt}}
\multiput(413.00,608.93)(1.033,-0.482){9}{\rule{0.900pt}{0.116pt}}
\multiput(413.00,609.17)(10.132,-6.000){2}{\rule{0.450pt}{0.400pt}}
\multiput(425.00,602.93)(0.943,-0.482){9}{\rule{0.833pt}{0.116pt}}
\multiput(425.00,603.17)(9.270,-6.000){2}{\rule{0.417pt}{0.400pt}}
\multiput(436.00,596.93)(1.267,-0.477){7}{\rule{1.060pt}{0.115pt}}
\multiput(436.00,597.17)(9.800,-5.000){2}{\rule{0.530pt}{0.400pt}}
\multiput(448.00,591.93)(1.033,-0.482){9}{\rule{0.900pt}{0.116pt}}
\multiput(448.00,592.17)(10.132,-6.000){2}{\rule{0.450pt}{0.400pt}}
\multiput(460.00,585.93)(1.155,-0.477){7}{\rule{0.980pt}{0.115pt}}
\multiput(460.00,586.17)(8.966,-5.000){2}{\rule{0.490pt}{0.400pt}}
\multiput(471.00,580.93)(1.033,-0.482){9}{\rule{0.900pt}{0.116pt}}
\multiput(471.00,581.17)(10.132,-6.000){2}{\rule{0.450pt}{0.400pt}}
\multiput(483.00,574.93)(1.155,-0.477){7}{\rule{0.980pt}{0.115pt}}
\multiput(483.00,575.17)(8.966,-5.000){2}{\rule{0.490pt}{0.400pt}}
\multiput(494.00,569.93)(1.267,-0.477){7}{\rule{1.060pt}{0.115pt}}
\multiput(494.00,570.17)(9.800,-5.000){2}{\rule{0.530pt}{0.400pt}}
\multiput(506.00,564.93)(1.033,-0.482){9}{\rule{0.900pt}{0.116pt}}
\multiput(506.00,565.17)(10.132,-6.000){2}{\rule{0.450pt}{0.400pt}}
\multiput(518.00,558.93)(1.155,-0.477){7}{\rule{0.980pt}{0.115pt}}
\multiput(518.00,559.17)(8.966,-5.000){2}{\rule{0.490pt}{0.400pt}}
\multiput(529.00,553.93)(1.267,-0.477){7}{\rule{1.060pt}{0.115pt}}
\multiput(529.00,554.17)(9.800,-5.000){2}{\rule{0.530pt}{0.400pt}}
\multiput(541.00,548.93)(1.155,-0.477){7}{\rule{0.980pt}{0.115pt}}
\multiput(541.00,549.17)(8.966,-5.000){2}{\rule{0.490pt}{0.400pt}}
\multiput(552.00,543.93)(1.267,-0.477){7}{\rule{1.060pt}{0.115pt}}
\multiput(552.00,544.17)(9.800,-5.000){2}{\rule{0.530pt}{0.400pt}}
\multiput(564.00,538.93)(1.267,-0.477){7}{\rule{1.060pt}{0.115pt}}
\multiput(564.00,539.17)(9.800,-5.000){2}{\rule{0.530pt}{0.400pt}}
\multiput(576.00,533.93)(1.155,-0.477){7}{\rule{0.980pt}{0.115pt}}
\multiput(576.00,534.17)(8.966,-5.000){2}{\rule{0.490pt}{0.400pt}}
\multiput(587.00,528.94)(1.651,-0.468){5}{\rule{1.300pt}{0.113pt}}
\multiput(587.00,529.17)(9.302,-4.000){2}{\rule{0.650pt}{0.400pt}}
\multiput(599.00,524.93)(1.267,-0.477){7}{\rule{1.060pt}{0.115pt}}
\multiput(599.00,525.17)(9.800,-5.000){2}{\rule{0.530pt}{0.400pt}}
\multiput(611.00,519.93)(1.155,-0.477){7}{\rule{0.980pt}{0.115pt}}
\multiput(611.00,520.17)(8.966,-5.000){2}{\rule{0.490pt}{0.400pt}}
\multiput(622.00,514.93)(1.267,-0.477){7}{\rule{1.060pt}{0.115pt}}
\multiput(622.00,515.17)(9.800,-5.000){2}{\rule{0.530pt}{0.400pt}}
\multiput(634.00,509.94)(1.505,-0.468){5}{\rule{1.200pt}{0.113pt}}
\multiput(634.00,510.17)(8.509,-4.000){2}{\rule{0.600pt}{0.400pt}}
\multiput(645.00,505.93)(1.267,-0.477){7}{\rule{1.060pt}{0.115pt}}
\multiput(645.00,506.17)(9.800,-5.000){2}{\rule{0.530pt}{0.400pt}}
\multiput(657.00,500.94)(1.651,-0.468){5}{\rule{1.300pt}{0.113pt}}
\multiput(657.00,501.17)(9.302,-4.000){2}{\rule{0.650pt}{0.400pt}}
\multiput(669.00,496.93)(1.155,-0.477){7}{\rule{0.980pt}{0.115pt}}
\multiput(669.00,497.17)(8.966,-5.000){2}{\rule{0.490pt}{0.400pt}}
\multiput(680.00,491.94)(1.651,-0.468){5}{\rule{1.300pt}{0.113pt}}
\multiput(680.00,492.17)(9.302,-4.000){2}{\rule{0.650pt}{0.400pt}}
\multiput(692.00,487.94)(1.505,-0.468){5}{\rule{1.200pt}{0.113pt}}
\multiput(692.00,488.17)(8.509,-4.000){2}{\rule{0.600pt}{0.400pt}}
\multiput(703.00,483.93)(1.267,-0.477){7}{\rule{1.060pt}{0.115pt}}
\multiput(703.00,484.17)(9.800,-5.000){2}{\rule{0.530pt}{0.400pt}}
\multiput(715.00,478.94)(1.651,-0.468){5}{\rule{1.300pt}{0.113pt}}
\multiput(715.00,479.17)(9.302,-4.000){2}{\rule{0.650pt}{0.400pt}}
\multiput(727.00,474.94)(1.505,-0.468){5}{\rule{1.200pt}{0.113pt}}
\multiput(727.00,475.17)(8.509,-4.000){2}{\rule{0.600pt}{0.400pt}}
\multiput(738.00,470.94)(1.651,-0.468){5}{\rule{1.300pt}{0.113pt}}
\multiput(738.00,471.17)(9.302,-4.000){2}{\rule{0.650pt}{0.400pt}}
\multiput(750.00,466.94)(1.505,-0.468){5}{\rule{1.200pt}{0.113pt}}
\multiput(750.00,467.17)(8.509,-4.000){2}{\rule{0.600pt}{0.400pt}}
\multiput(761.00,462.94)(1.651,-0.468){5}{\rule{1.300pt}{0.113pt}}
\multiput(761.00,463.17)(9.302,-4.000){2}{\rule{0.650pt}{0.400pt}}
\multiput(773.00,458.94)(1.651,-0.468){5}{\rule{1.300pt}{0.113pt}}
\multiput(773.00,459.17)(9.302,-4.000){2}{\rule{0.650pt}{0.400pt}}
\multiput(785.00,454.94)(1.505,-0.468){5}{\rule{1.200pt}{0.113pt}}
\multiput(785.00,455.17)(8.509,-4.000){2}{\rule{0.600pt}{0.400pt}}
\multiput(796.00,450.94)(1.651,-0.468){5}{\rule{1.300pt}{0.113pt}}
\multiput(796.00,451.17)(9.302,-4.000){2}{\rule{0.650pt}{0.400pt}}
\multiput(808.00,446.94)(1.505,-0.468){5}{\rule{1.200pt}{0.113pt}}
\multiput(808.00,447.17)(8.509,-4.000){2}{\rule{0.600pt}{0.400pt}}
\multiput(819.00,442.94)(1.651,-0.468){5}{\rule{1.300pt}{0.113pt}}
\multiput(819.00,443.17)(9.302,-4.000){2}{\rule{0.650pt}{0.400pt}}
\multiput(831.00,438.94)(1.651,-0.468){5}{\rule{1.300pt}{0.113pt}}
\multiput(831.00,439.17)(9.302,-4.000){2}{\rule{0.650pt}{0.400pt}}
\multiput(843.00,434.94)(1.505,-0.468){5}{\rule{1.200pt}{0.113pt}}
\multiput(843.00,435.17)(8.509,-4.000){2}{\rule{0.600pt}{0.400pt}}
\multiput(854.00,430.95)(2.472,-0.447){3}{\rule{1.700pt}{0.108pt}}
\multiput(854.00,431.17)(8.472,-3.000){2}{\rule{0.850pt}{0.400pt}}
\multiput(866.00,427.94)(1.651,-0.468){5}{\rule{1.300pt}{0.113pt}}
\multiput(866.00,428.17)(9.302,-4.000){2}{\rule{0.650pt}{0.400pt}}
\multiput(878.00,423.94)(1.505,-0.468){5}{\rule{1.200pt}{0.113pt}}
\multiput(878.00,424.17)(8.509,-4.000){2}{\rule{0.600pt}{0.400pt}}
\multiput(889.00,419.95)(2.472,-0.447){3}{\rule{1.700pt}{0.108pt}}
\multiput(889.00,420.17)(8.472,-3.000){2}{\rule{0.850pt}{0.400pt}}
\multiput(901.00,416.94)(1.505,-0.468){5}{\rule{1.200pt}{0.113pt}}
\multiput(901.00,417.17)(8.509,-4.000){2}{\rule{0.600pt}{0.400pt}}
\multiput(912.00,412.95)(2.472,-0.447){3}{\rule{1.700pt}{0.108pt}}
\multiput(912.00,413.17)(8.472,-3.000){2}{\rule{0.850pt}{0.400pt}}
\multiput(924.00,409.94)(1.651,-0.468){5}{\rule{1.300pt}{0.113pt}}
\multiput(924.00,410.17)(9.302,-4.000){2}{\rule{0.650pt}{0.400pt}}
\multiput(936.00,405.95)(2.248,-0.447){3}{\rule{1.567pt}{0.108pt}}
\multiput(936.00,406.17)(7.748,-3.000){2}{\rule{0.783pt}{0.400pt}}
\multiput(947.00,402.94)(1.651,-0.468){5}{\rule{1.300pt}{0.113pt}}
\multiput(947.00,403.17)(9.302,-4.000){2}{\rule{0.650pt}{0.400pt}}
\multiput(959.00,398.95)(2.248,-0.447){3}{\rule{1.567pt}{0.108pt}}
\multiput(959.00,399.17)(7.748,-3.000){2}{\rule{0.783pt}{0.400pt}}
\multiput(970.00,395.94)(1.651,-0.468){5}{\rule{1.300pt}{0.113pt}}
\multiput(970.00,396.17)(9.302,-4.000){2}{\rule{0.650pt}{0.400pt}}
\multiput(982.00,391.95)(2.472,-0.447){3}{\rule{1.700pt}{0.108pt}}
\multiput(982.00,392.17)(8.472,-3.000){2}{\rule{0.850pt}{0.400pt}}
\multiput(994.00,388.95)(2.248,-0.447){3}{\rule{1.567pt}{0.108pt}}
\multiput(994.00,389.17)(7.748,-3.000){2}{\rule{0.783pt}{0.400pt}}
\multiput(1005.00,385.95)(2.472,-0.447){3}{\rule{1.700pt}{0.108pt}}
\multiput(1005.00,386.17)(8.472,-3.000){2}{\rule{0.850pt}{0.400pt}}
\multiput(1017.00,382.94)(1.505,-0.468){5}{\rule{1.200pt}{0.113pt}}
\multiput(1017.00,383.17)(8.509,-4.000){2}{\rule{0.600pt}{0.400pt}}
\multiput(1028.00,378.95)(2.472,-0.447){3}{\rule{1.700pt}{0.108pt}}
\multiput(1028.00,379.17)(8.472,-3.000){2}{\rule{0.850pt}{0.400pt}}
\multiput(1040.00,375.95)(2.472,-0.447){3}{\rule{1.700pt}{0.108pt}}
\multiput(1040.00,376.17)(8.472,-3.000){2}{\rule{0.850pt}{0.400pt}}
\multiput(1052.00,372.95)(2.248,-0.447){3}{\rule{1.567pt}{0.108pt}}
\multiput(1052.00,373.17)(7.748,-3.000){2}{\rule{0.783pt}{0.400pt}}
\multiput(1063.00,369.95)(2.472,-0.447){3}{\rule{1.700pt}{0.108pt}}
\multiput(1063.00,370.17)(8.472,-3.000){2}{\rule{0.850pt}{0.400pt}}
\multiput(1075.00,366.95)(2.472,-0.447){3}{\rule{1.700pt}{0.108pt}}
\multiput(1075.00,367.17)(8.472,-3.000){2}{\rule{0.850pt}{0.400pt}}
\multiput(1087.00,363.95)(2.248,-0.447){3}{\rule{1.567pt}{0.108pt}}
\multiput(1087.00,364.17)(7.748,-3.000){2}{\rule{0.783pt}{0.400pt}}
\multiput(1098.00,360.95)(2.472,-0.447){3}{\rule{1.700pt}{0.108pt}}
\multiput(1098.00,361.17)(8.472,-3.000){2}{\rule{0.850pt}{0.400pt}}
\multiput(1110.00,357.95)(2.248,-0.447){3}{\rule{1.567pt}{0.108pt}}
\multiput(1110.00,358.17)(7.748,-3.000){2}{\rule{0.783pt}{0.400pt}}
\multiput(1121.00,354.95)(2.472,-0.447){3}{\rule{1.700pt}{0.108pt}}
\multiput(1121.00,355.17)(8.472,-3.000){2}{\rule{0.850pt}{0.400pt}}
\multiput(1133.00,351.95)(2.472,-0.447){3}{\rule{1.700pt}{0.108pt}}
\multiput(1133.00,352.17)(8.472,-3.000){2}{\rule{0.850pt}{0.400pt}}
\multiput(1145.00,348.95)(2.248,-0.447){3}{\rule{1.567pt}{0.108pt}}
\multiput(1145.00,349.17)(7.748,-3.000){2}{\rule{0.783pt}{0.400pt}}
\multiput(1156.00,345.95)(2.472,-0.447){3}{\rule{1.700pt}{0.108pt}}
\multiput(1156.00,346.17)(8.472,-3.000){2}{\rule{0.850pt}{0.400pt}}
\multiput(1168.00,342.95)(2.248,-0.447){3}{\rule{1.567pt}{0.108pt}}
\multiput(1168.00,343.17)(7.748,-3.000){2}{\rule{0.783pt}{0.400pt}}
\multiput(1179.00,339.95)(2.472,-0.447){3}{\rule{1.700pt}{0.108pt}}
\multiput(1179.00,340.17)(8.472,-3.000){2}{\rule{0.850pt}{0.400pt}}
\put(1191,336.17){\rule{2.500pt}{0.400pt}}
\multiput(1191.00,337.17)(6.811,-2.000){2}{\rule{1.250pt}{0.400pt}}
\multiput(1203.00,334.95)(2.248,-0.447){3}{\rule{1.567pt}{0.108pt}}
\multiput(1203.00,335.17)(7.748,-3.000){2}{\rule{0.783pt}{0.400pt}}
\multiput(1214.00,331.95)(2.472,-0.447){3}{\rule{1.700pt}{0.108pt}}
\multiput(1214.00,332.17)(8.472,-3.000){2}{\rule{0.850pt}{0.400pt}}
\multiput(1226.00,328.95)(2.248,-0.447){3}{\rule{1.567pt}{0.108pt}}
\multiput(1226.00,329.17)(7.748,-3.000){2}{\rule{0.783pt}{0.400pt}}
\put(1237,325.17){\rule{2.500pt}{0.400pt}}
\multiput(1237.00,326.17)(6.811,-2.000){2}{\rule{1.250pt}{0.400pt}}
\multiput(1249.00,323.95)(2.472,-0.447){3}{\rule{1.700pt}{0.108pt}}
\multiput(1249.00,324.17)(8.472,-3.000){2}{\rule{0.850pt}{0.400pt}}
\multiput(1261.00,320.95)(2.248,-0.447){3}{\rule{1.567pt}{0.108pt}}
\multiput(1261.00,321.17)(7.748,-3.000){2}{\rule{0.783pt}{0.400pt}}
\put(1272,317.17){\rule{2.500pt}{0.400pt}}
\multiput(1272.00,318.17)(6.811,-2.000){2}{\rule{1.250pt}{0.400pt}}
\multiput(1284.00,315.95)(2.248,-0.447){3}{\rule{1.567pt}{0.108pt}}
\multiput(1284.00,316.17)(7.748,-3.000){2}{\rule{0.783pt}{0.400pt}}
\put(1295,312.17){\rule{2.500pt}{0.400pt}}
\multiput(1295.00,313.17)(6.811,-2.000){2}{\rule{1.250pt}{0.400pt}}
\multiput(1307.00,310.95)(2.472,-0.447){3}{\rule{1.700pt}{0.108pt}}
\multiput(1307.00,311.17)(8.472,-3.000){2}{\rule{0.850pt}{0.400pt}}
\put(1319,307.17){\rule{2.300pt}{0.400pt}}
\multiput(1319.00,308.17)(6.226,-2.000){2}{\rule{1.150pt}{0.400pt}}
\sbox{\plotpoint}{\rule[-0.500pt]{1.000pt}{1.000pt}}%
\put(181.00,758.00){\usebox{\plotpoint}}
\put(198.55,746.96){\usebox{\plotpoint}}
\put(216.11,735.93){\usebox{\plotpoint}}
\put(233.77,725.05){\usebox{\plotpoint}}
\put(251.25,713.86){\usebox{\plotpoint}}
\put(269.35,703.71){\usebox{\plotpoint}}
\put(287.01,692.82){\usebox{\plotpoint}}
\put(305.22,682.89){\usebox{\plotpoint}}
\put(323.13,672.44){\usebox{\plotpoint}}
\put(341.36,662.54){\usebox{\plotpoint}}
\put(359.62,652.69){\usebox{\plotpoint}}
\put(377.98,643.01){\usebox{\plotpoint}}
\put(396.54,633.73){\usebox{\plotpoint}}
\put(414.90,624.05){\usebox{\plotpoint}}
\put(433.31,614.47){\usebox{\plotpoint}}
\put(451.95,605.36){\usebox{\plotpoint}}
\put(470.56,596.24){\usebox{\plotpoint}}
\put(489.37,587.53){\usebox{\plotpoint}}
\put(508.29,579.05){\usebox{\plotpoint}}
\put(526.99,570.10){\usebox{\plotpoint}}
\put(545.97,561.74){\usebox{\plotpoint}}
\put(565.05,553.56){\usebox{\plotpoint}}
\put(584.09,545.32){\usebox{\plotpoint}}
\put(603.21,537.25){\usebox{\plotpoint}}
\put(622.57,529.76){\usebox{\plotpoint}}
\put(641.62,521.54){\usebox{\plotpoint}}
\put(661.05,514.31){\usebox{\plotpoint}}
\put(680.06,505.98){\usebox{\plotpoint}}
\put(699.68,499.21){\usebox{\plotpoint}}
\put(719.01,491.67){\usebox{\plotpoint}}
\put(738.58,484.76){\usebox{\plotpoint}}
\put(757.87,477.14){\usebox{\plotpoint}}
\put(777.54,470.49){\usebox{\plotpoint}}
\put(797.12,463.63){\usebox{\plotpoint}}
\put(816.73,456.83){\usebox{\plotpoint}}
\put(836.40,450.20){\usebox{\plotpoint}}
\put(856.03,443.49){\usebox{\plotpoint}}
\put(875.94,437.69){\usebox{\plotpoint}}
\put(895.53,430.82){\usebox{\plotpoint}}
\put(915.40,424.87){\usebox{\plotpoint}}
\put(935.34,419.16){\usebox{\plotpoint}}
\put(955.12,412.97){\usebox{\plotpoint}}
\put(974.90,406.77){\usebox{\plotpoint}}
\put(994.76,400.79){\usebox{\plotpoint}}
\put(1014.84,395.54){\usebox{\plotpoint}}
\put(1034.62,389.34){\usebox{\plotpoint}}
\put(1054.74,384.25){\usebox{\plotpoint}}
\put(1074.83,379.04){\usebox{\plotpoint}}
\put(1094.65,372.91){\usebox{\plotpoint}}
\put(1114.74,367.71){\usebox{\plotpoint}}
\put(1134.85,362.54){\usebox{\plotpoint}}
\put(1154.93,357.29){\usebox{\plotpoint}}
\put(1175.02,352.09){\usebox{\plotpoint}}
\put(1195.33,347.92){\usebox{\plotpoint}}
\put(1215.40,342.65){\usebox{\plotpoint}}
\put(1235.49,337.41){\usebox{\plotpoint}}
\put(1255.81,333.30){\usebox{\plotpoint}}
\put(1275.95,328.34){\usebox{\plotpoint}}
\put(1296.16,323.71){\usebox{\plotpoint}}
\put(1316.45,319.43){\usebox{\plotpoint}}
\put(1330,316){\usebox{\plotpoint}}
\put(181.00,723.00){\usebox{\plotpoint}}
\put(198.79,712.32){\usebox{\plotpoint}}
\put(216.58,701.63){\usebox{\plotpoint}}
\put(234.26,690.77){\usebox{\plotpoint}}
\put(252.16,680.26){\usebox{\plotpoint}}
\put(270.13,669.93){\usebox{\plotpoint}}
\put(288.03,659.48){\usebox{\plotpoint}}
\put(306.27,649.59){\usebox{\plotpoint}}
\put(324.53,639.73){\usebox{\plotpoint}}
\put(343.09,630.45){\usebox{\plotpoint}}
\put(361.45,620.77){\usebox{\plotpoint}}
\put(379.81,611.10){\usebox{\plotpoint}}
\put(398.37,601.81){\usebox{\plotpoint}}
\put(416.85,592.40){\usebox{\plotpoint}}
\put(435.47,583.29){\usebox{\plotpoint}}
\put(454.40,574.80){\usebox{\plotpoint}}
\put(473.22,566.07){\usebox{\plotpoint}}
\put(491.92,557.13){\usebox{\plotpoint}}
\put(510.97,548.93){\usebox{\plotpoint}}
\put(529.98,540.59){\usebox{\plotpoint}}
\put(549.03,532.35){\usebox{\plotpoint}}
\put(568.14,524.27){\usebox{\plotpoint}}
\put(587.50,516.79){\usebox{\plotpoint}}
\put(606.66,508.81){\usebox{\plotpoint}}
\put(626.01,501.33){\usebox{\plotpoint}}
\put(645.02,492.99){\usebox{\plotpoint}}
\put(664.71,486.43){\usebox{\plotpoint}}
\put(683.94,478.69){\usebox{\plotpoint}}
\put(703.52,471.83){\usebox{\plotpoint}}
\put(722.99,464.67){\usebox{\plotpoint}}
\put(742.47,457.51){\usebox{\plotpoint}}
\put(762.05,450.65){\usebox{\plotpoint}}
\put(781.74,444.09){\usebox{\plotpoint}}
\put(801.33,437.22){\usebox{\plotpoint}}
\put(821.20,431.27){\usebox{\plotpoint}}
\put(840.89,424.70){\usebox{\plotpoint}}
\put(860.63,418.34){\usebox{\plotpoint}}
\put(880.48,412.32){\usebox{\plotpoint}}
\put(900.31,406.23){\usebox{\plotpoint}}
\put(920.07,399.98){\usebox{\plotpoint}}
\put(940.08,394.52){\usebox{\plotpoint}}
\put(959.96,388.65){\usebox{\plotpoint}}
\put(979.77,382.56){\usebox{\plotpoint}}
\put(999.87,377.40){\usebox{\plotpoint}}
\put(1019.70,371.26){\usebox{\plotpoint}}
\put(1039.78,366.05){\usebox{\plotpoint}}
\put(1059.88,360.85){\usebox{\plotpoint}}
\put(1079.99,355.75){\usebox{\plotpoint}}
\put(1100.07,350.48){\usebox{\plotpoint}}
\put(1120.15,345.23){\usebox{\plotpoint}}
\put(1140.28,340.18){\usebox{\plotpoint}}
\put(1160.43,335.26){\usebox{\plotpoint}}
\put(1180.63,330.59){\usebox{\plotpoint}}
\put(1200.76,325.56){\usebox{\plotpoint}}
\put(1221.05,321.24){\usebox{\plotpoint}}
\put(1241.19,316.30){\usebox{\plotpoint}}
\put(1261.47,311.92){\usebox{\plotpoint}}
\put(1281.75,307.56){\usebox{\plotpoint}}
\put(1301.94,302.84){\usebox{\plotpoint}}
\put(1322.20,298.42){\usebox{\plotpoint}}
\put(1330,297){\usebox{\plotpoint}}
\end{picture}
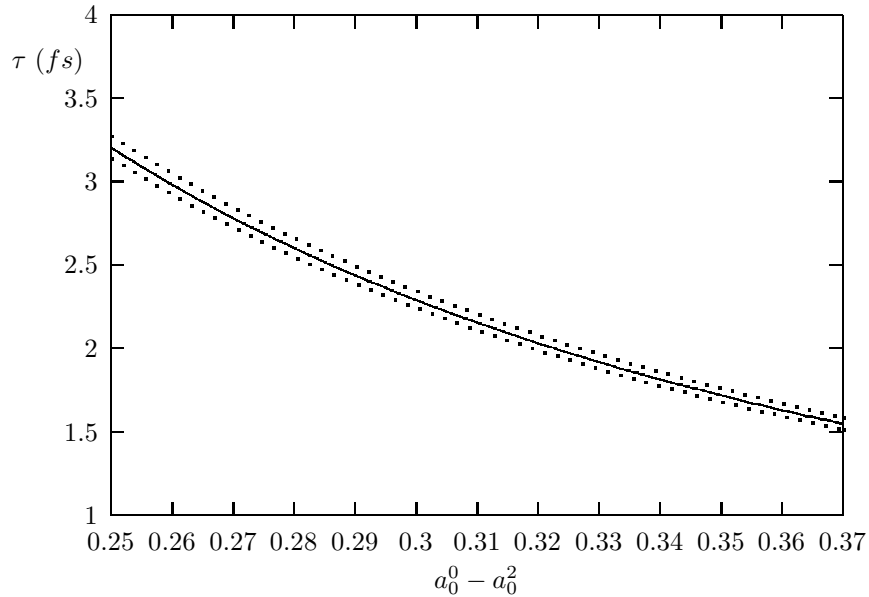
\caption{The pionium lifetime as a function of the combination
$(a_0^0-a_0^2)$ of the $S$-wave scattering lengths (full line). 
The band delineated by the dotted lines takes into account the estimated
uncertainties (2-2.5\%).}
\lb{f1}
\end{center}
\efg
Values of the lifetime close to 3 fs, lying above 2.9 fs, say, would 
confirm the scheme of standard $\chi PT$. Values of the lifetime lying 
below 2.4 fs remain outside the domain of predictions of standard $\chi PT$
and would necessitate an alternative scheme of chiral symmetry breaking.
Values of the lifetime lying in the interval 2.4-2.9 fs, because of the
possibly existing uncertainties, would be more difficult to interpret and
would require a more refined analysis.
\par

\end{document}